\documentstyle[twocolumn,epsfig]{jpsjnew}

\newcommand{\vereq}[2]{\lower3pt\vbox{\baselineskip1.5pt \lineskip1.5pt
\ialign{$#1\hfill##\hfil$\crcr#2\crcr\sim\crcr}}}

\newcommand{\gtrsim}{\mathrel{\mathpalette\vereq>}}

\newcommand{\agt}{\gtrsim}

\def\runtitle{
Stability Analysis of Optimal Velocity Model for Traffic and Granular
Flow under Open Boundary Condition}
\def\runauthor{Namiko {\sc Mitarai} and Hiizu {\sc Nakanishi}}

\title
{
Stability Analysis of Optimal Velocity Model for Traffic and Granular
Flow under Open Boundary Condition
}

\author
{ 
Namiko {\sc Mitarai}\footnote{E-mail: namiko@stat.phys.kyushu-u.ac.jp} 
and Hiizu {\sc Nakanishi}\footnote{E-mail: naka4scp@mbox.nc.kyushu-u.ac.jp} 
}

\inst
{
Department of Physics, Kyushu University 33, Fukuoka 812-8581
}

\abst
{
We analyzed the stability of the uniform flow solution in the optimal
velocity model for traffic and granular flow under the open
boundary condition.
It was demonstrated that, even within the linearly unstable region, there is a
parameter region where the uniform solution is stable against a
localized perturbation.
We also found an oscillatory solution in the linearly unstable region
and its period is not commensurate with the periodicity of the 
car index space.
The oscillatory solution has some features in common with the
synchronized flow observed in real traffic.
}

\kword
{
optimal velocity model,
traffic flow, granular pipe flow, open boundary condition,
stability of uniform solution
}

\begin{document}
\sloppy
\maketitle

The traffic flow on a highway~\cite{MH,KR1,KR2,KR3,K}  and
the granular pipe flow~\cite{HNNM,HINNM,MNMH,MIKKRM,ASRH}
are similar in a number of characteristics.
Both consist of discrete elements following dissipative dynamics,
both are quasi one-dimensional systems, 
and spontaneous density waves with the power law occur in both systems.
To understand the common structure of such phenomena,
both discrete and continuous models have been suggested and
have succeeded in reproducing some aspects of the spatiotemporal
patterns.~\cite{MNMH,BHNSS1,BHNSS2,KS,PH1,PH2,KK1,KK2}

In recent years, a car-following model, termed the 
``optimal velocity model'' (OV model),
was proposed.~\cite{BHNSS1,BHNSS2}
The model is realistic enough to be able to reproduce a spontaneous
traffic jam as in real traffic, but simple enough to 
be regarded as a model of granular pipe flow without
much modification.
In fact, a model similar to the OV model has been proposed
as a granular flow model
and it has been demonstrated to exhibit
power law behavior in the power spectra of the 
density wave in certain situations.~\cite{MNMH}

Even though a number of analyses of the OV model have been performed,
most of them were conducted under the periodic boundary condition 
(PBC).~\cite{BHNSS1,BHNSS2,KS}
The PBC corresponds to the situation that cars run in a circuit,
but it is not suitable for most situations in real traffic observations.
When we consider the situation of granular pipe flow,
the PBC is quite unrealistic.
Furthermore, from the experimental results of the granular pipe flow,
it has been suggested that the condition of the downstream boundary is
important.~\cite{MIKKRM}

In the present work,
we investigate the OV model under the open boundary condition (OBC),
and examine the effects of the boundary condition on
the stability of the system.

%------figure 1------------------------------------------
\begin{figure}
\begin{center}
\epsfig{file=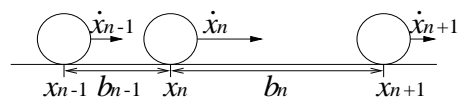,width=6cm,angle=-90,clip=}
\end{center}
\caption{
The optimal velocity model.
}
\label{OVmodel}
\end{figure}
%---------end of figure 1------------------------------
The OV model is a dynamical model for a one-dimensional system which
consists of discrete elements; the elements can be cars or grain
particles, but they tend to run at the speed determined by a local
configuration.  When the ($n+1$)th car precedes the $n$th car
(Fig. \ref{OVmodel}) and 
the position of the $n$th car at the time $t$ is denoted by $x_n(t)$, 
then the dynamics is governed by the equation of motion
\begin{equation}
   \ddot x_n(t)=a\bigl[ U(b_n(t))-\dot x_n(t)\bigr] , 
\label{eq:OVM}
\end{equation}
where $b_n(t)$ represents the headway of the $n$th car defined by
\begin{equation}
  b_n(t)=x_{n+1}(t)-x_n(t),
\label{eq:b_j}
\end{equation}
and the dot denotes the time derivative.
Here, $a$ is a sensitivity constant and
$U(b)$ is the optimal velocity that the drivers prefer when 
the headway is $b$.
For $U(b)$, we employ
\begin{equation}
U(b)=\tanh (b-2)+\tanh(2),
\end{equation}
as in most other works on the OV model.

Equation (\ref{eq:OVM}) has a uniform solution
\begin{equation}
x_n(t)=\bar{b}n+U(\bar{b})t+\mbox{const.},
\label{eq:uniform}
\end{equation}
which represents that all of the cars run with the same headway $\bar{b}$ and
the same optimal velocity $U(\bar{b})$.
It is easy to show~\cite{BHNSS1,BHNSS2} that 
the solution is linearly unstable when
\begin{equation}
a<2U'(\bar{b}),
\label{eq:unstable}
\end{equation}
where the prime represents a derivative.
The behavior of the model in the linearly unstable region has been
studied and demonstrated~\cite{BHNSS1,BHNSS2,KS} that 
the system eventually segregates into two
regions, namely, the jammed region and the free-flow region.
The headways in the jammed and the free-flow region are determined by 
$a$ for a given OV function $U(b)$.~\cite{BHNSS2,KS}

For the purpose of analyzing the system behavior in the situation where the
upper and lower streams are distinguished, we employ the OBC, and define it 
as follows:
at $x=0$, the cars enter the system at a constant time
interval $\bar b/U(\bar b)$ and velocity $U(\bar b)$, and 
the car that is farthest ahead, 
which has the largest index and no car to follow
within the system, 
travels as if its headway is $\bar b$, namely, it 
follows the equation of motion as
\begin{equation}
\ddot{x}_{f}=a[U(\bar{b})-\dot{x}_{f}],
\end{equation}
until it leaves the system at $x=L$.

In order to examine the stability,
we perturb the system at $t=0$ locally in space and time by shifting
the velocity of the car with $n=0$ at $x=L/2$;
in the actual simulations, the system is
prepared with the initial condition as
\begin{equation}
x_n(0) = \bar b n+L/2,
\quad
\dot x_n(0) = U(\bar b)
\quad
\mbox{ for } n = \pm 1, \pm 2, ...,
\end{equation}
\begin{equation}
x_0(0) = L/2,
\quad
\dot x_0(0) = U(\bar b)+\epsilon ,
\end{equation}
with a small value of $\epsilon$.
%-------figure 2-------------------------------------------
\begin{figure}
\begin{center}
\epsfig{file=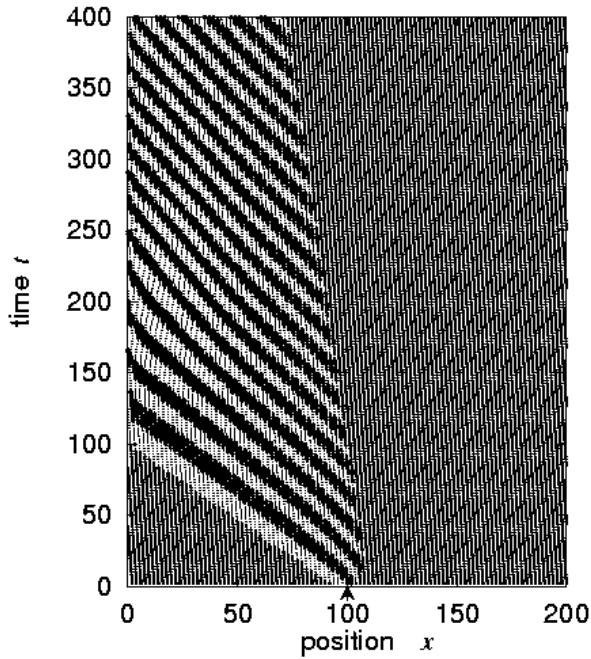, width=8cm, angle=-90}
\end{center}
\caption{
Spatiotemporal diagram of the OV model with $a=1.4$,
$\bar{b}=2$, $L=200$ and $\epsilon=0.1$. 
The system is perturbed at the point indicated by the arrow on the
bottom line.
Positions of cars are indicated by dots, whose 
size represents the headway of the car;
the large, medium-sized, and small dots represent
cars with headways smaller than, equal to, and larger than $\bar{b}$, 
respectively.
As a result, the original uniform flow region appears shaded, and the
region with $b_n>\bar b$ ($b_n<\bar b$) appears lighter (darker).
The disturbance recedes towards the upper stream.
}
\label{ova1c4b2}
\end{figure}
%------end of figure 2----------------------
% ------------------ Figure 3 ---------------
\begin{figure}
\begin{center}
\epsfig{file=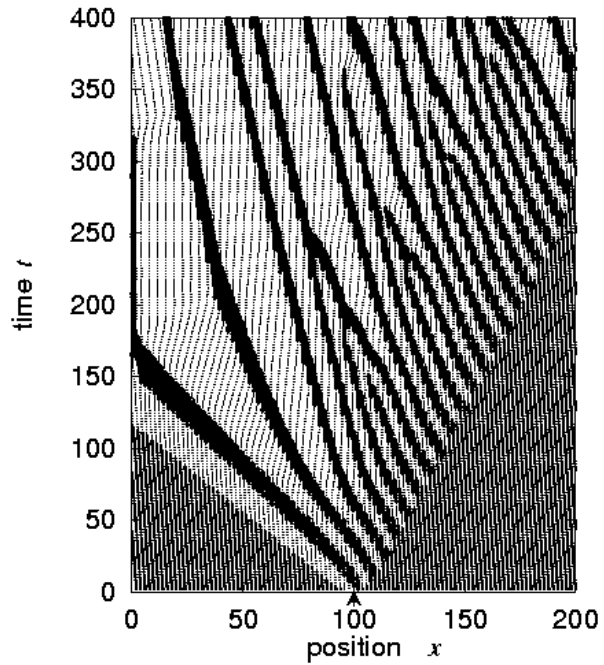, width=8cm, angle=-90}
\end{center}
\caption{
Spatiotemporal diagram  with $a=1$,
$\bar{b}=2$, $L=200$ and $\epsilon=0.1$. 
The disturbance spreads in both directions.
}
\label{ova1b2}
\end{figure}
%---------------end of figure 3--------------------------

\begin{figure}
\begin{center}
\epsfig{file=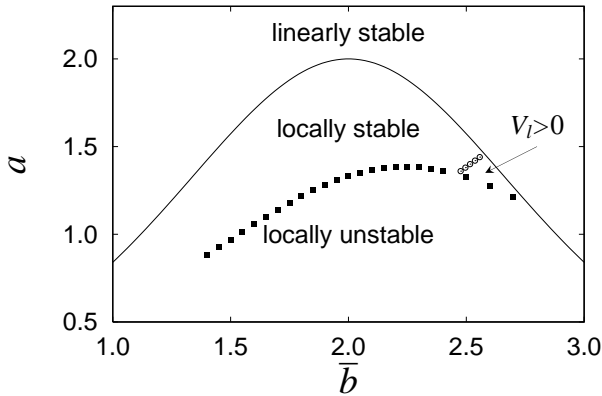,width=8cm,angle=-90}
\end{center}
\caption{
Phase diagram in the $\bar b-a$ plane.
The solid line represents the linear stability limit above which the
uniform solution is linearly stable.
The solid boxes represent the local stability limit determined by
$V_0=0$, and the open circles denote the boundary where $V_l=0$.
}
\label{phase}
\end{figure}

Figures \ref{ova1c4b2} and \ref{ova1b2} show two typical spatiotemporal diagrams, where the
original uniform region with the constant headway $b_n=\bar b$ is
denoted by the shaded region, and the region with $b_n>\bar b$ and that with
$b_n<\bar b$ are shown by the lighter and the darker region, respectively.
The perturbation is given at $x=L/2$ and $t=0$.
For both cases, the parameters are in the linearly unstable
region, thus, the largest response to the perturbation over the entire
system grows as time passes.
However, in the case of Fig. \ref{ova1c4b2} with the larger value of $a$, 
the effects of 
the perturbation travel only towards the upper stream and eventually
exit the system, then, the entire system recovers the uniform
configuration.
On the other hand, in the case of Fig. \ref{ova1b2} 
with the smaller value of $a$, the 
effects of the perturbation extend in both directions and the uniform
region is eliminated completely.
Therefore, within the linearly unstable region in the $\bar b-a$
plane, we can draw a line that separates the two distinct parameter regions:
{\it the locally stable region} where a small local disturbance in the
uniform state recedes, and {\it the locally unstable region} where the
local disturbance extends (Fig. \ref{phase}).~\cite{smalla}
%----------------figure 4---------------------------
%------------------end of figure 4-------------------

%---------------figure 5--------------------
\begin{figure}
\begin{center}
\begin{minipage}[h]{7cm}
\epsfig{file=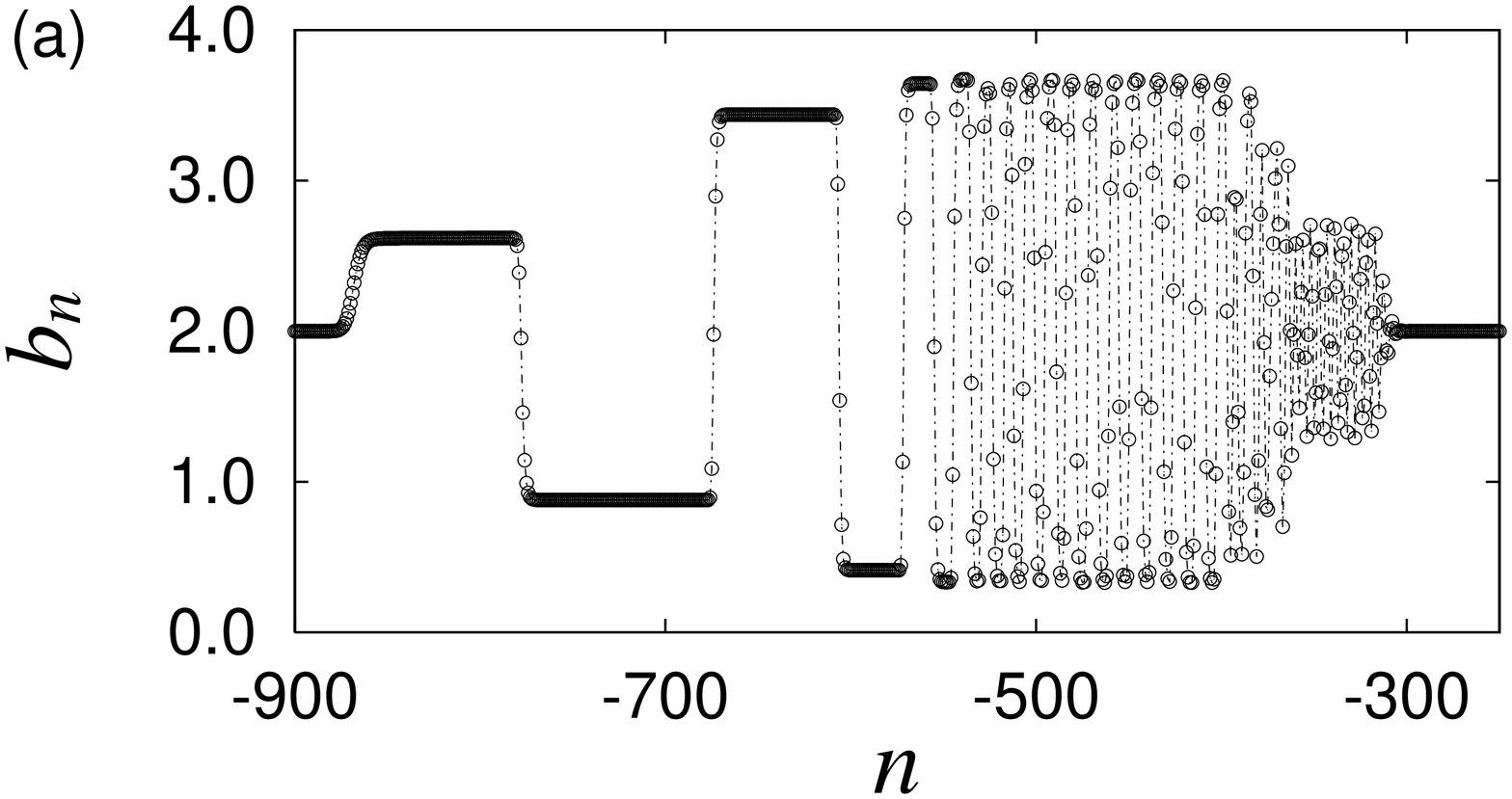, width=7.8cm,angle=-90}
\epsfig{file=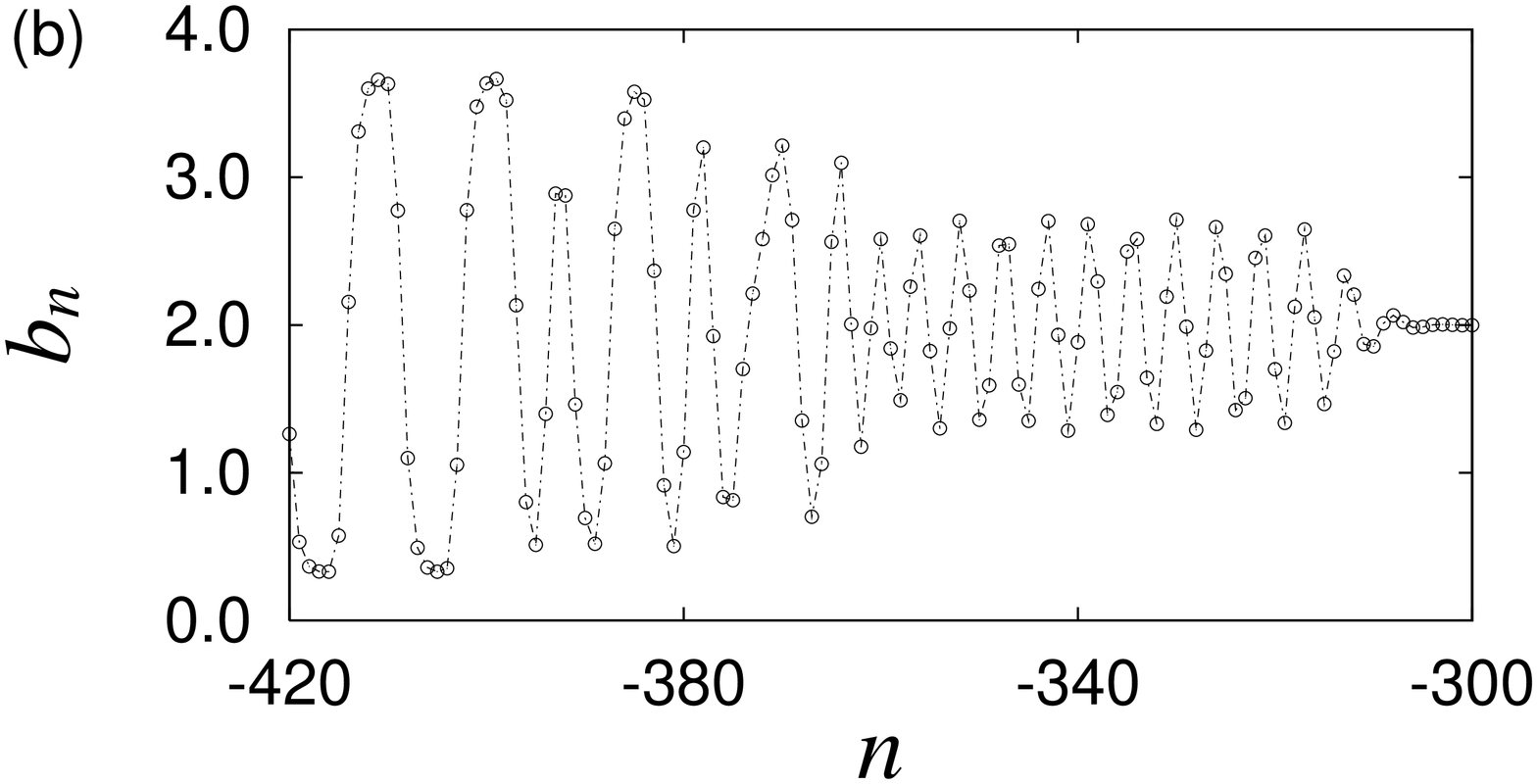, width=7.8cm,angle=-90}
\epsfig{file=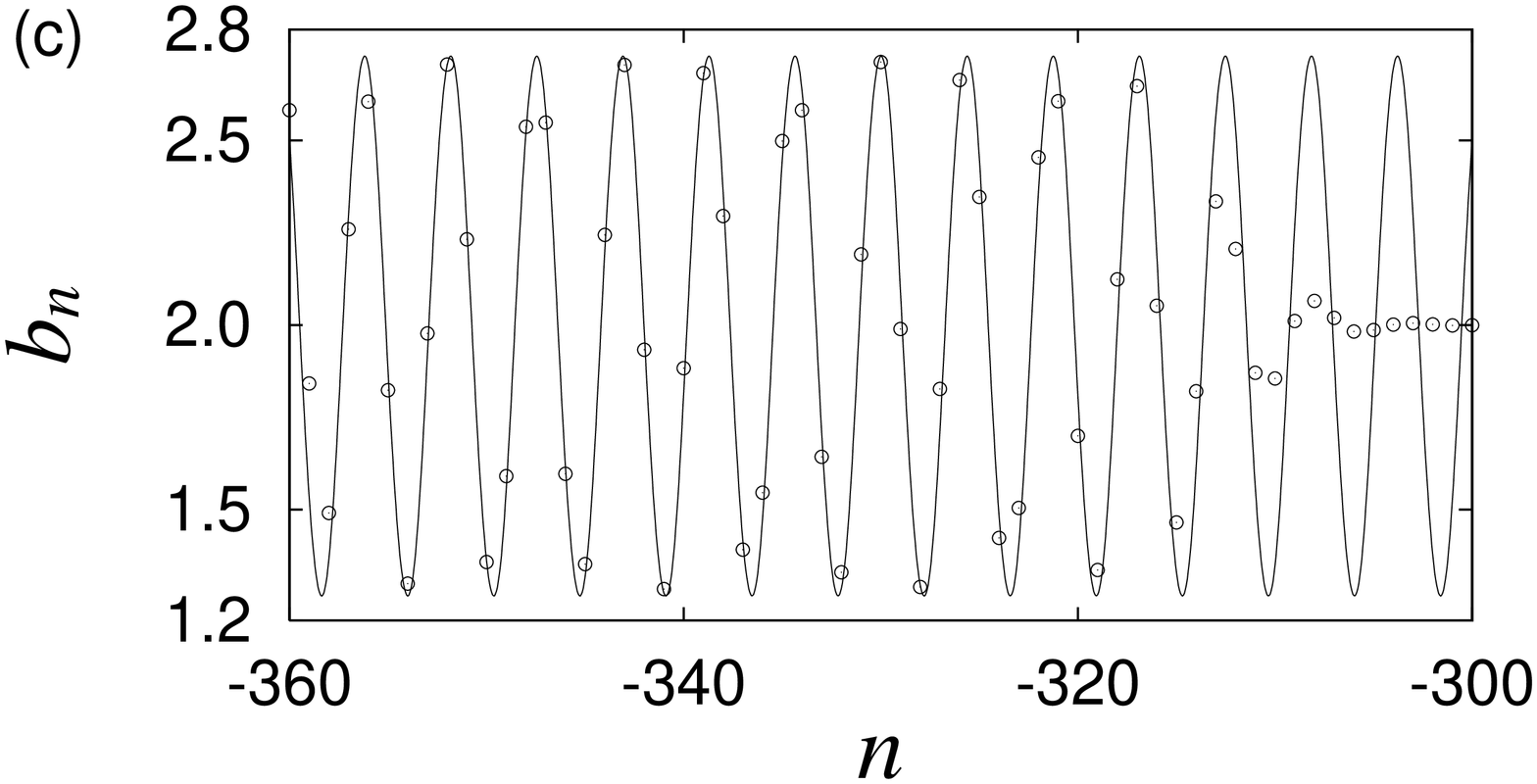, width=7.8cm,angle=-90}
\end{minipage}
\end{center}
\caption{
Snapshots of car configuration represented by 
the $b_n$ vs $n$ plot at $t=988$ with $a=1, \bar{b}=2, L=10000$
(the open circles connected by the dashed line).
(a) The oscillatory flow region appears around $-360<n<-300$ followed by
the alternating jammed and free-flow regions around
$-550<n<-400$.
(b) The magnification of (a). 
The oscillatory behavior can be seen clearly.
(c) The same data of $b_n$ are plotted with
the solid line which represents the result from eq. (\ref{eq:foscillate}) with
$c=c_s=0.61$. 
}
\label{oscillate}
\end{figure}
%---------end of figure 5-------------------------------
What is peculiar to these spatiotemporal diagrams is that there appears
a region with a regular striping pattern of lighter and darker regions
behind the uniform flow region.
The behavior of the system in this region can be seen more clearly in the
snapshots of the headway of the cars.
In Fig. \ref{oscillate}(a), 
the right end region at $n \agt -300$ represents the
uniform flow, and the fluctuating region around the part near
$n\approx -450$ represents an alternating sequence of jammed and free-flow
regions;
the upper and lower limiting values of $b_n$ in this region are close to 
those of the jammed and free-flow regions in the PBC.
Between the uniform and the alternating region, there exists {\it the
oscillatory flow} region which has a smaller amplitude of fluctuation than
that of the alternating region (Fig. \ref{oscillate}(b));
this oscillatory flow region corresponds to the regular stripes in
Fig. \ref{ova1b2}.

This oscillatory flow has some interesting characteristics, as
can be seen from Fig. \ref{oscillate}:
(i)
The oscillatory flow region appears directly behind the uniform flow, and
the oscillatory flow is followed by the alternating region of the free-flow
and the jammed region
when the system size $L$ is sufficiently large. 
(ii)
The amplitude of the oscillatory flow (the difference between the minimum and 
maximum value of $b_n$ in the region) is smaller than the difference 
in the headway between the jammed and the free-flow regions.
(iii)
The oscillatory flow region typically extends over a hundred cars and
persists.
(iv)
The velocity and the headway of the cars oscillate periodically 
around that of the uniform flow, but
their periodicity is incommensurate with 
the period of the car index space.

Because the oscillatory flow appears spontaneously 
being triggered by a local disturbance,
it is expected that there exists, at least, a metastable solution where the
oscillatory flow extends over the entire system.
In order to find this oscillatory solution,
we assume it is represented as
\begin{equation}
b_n=\bar{b}+f(n+ct),
\label{eq:oscillate}
\end{equation}
with the phase speed $c$;
$\bar{b}$ is the mean headway, therefore the function $f(n+ct)$
should have zero mean.
Substituting eq. (\ref{eq:oscillate}) into eq. (\ref{eq:OVM}), we obtain
\begin{equation}
c^2f''(z)=a[U(\bar{b}+f(z+1))-U(\bar{b}+f(z))-cf'(z)].
\label{eq:foscillate}
\end{equation}

This can be solved numerically and is found to have an oscillatory
solution for a finite range of the phase speed $c$;
{\it e.g.}, $c\in (0.56, 0.64)$ for $a=1$ and $\bar b=2$.
If we set the phase speed at the value obtained by the direct
simulation of eq. (\ref{eq:OVM}), namely, $c=c_s= 0.61$ for $a=1$ and
$\bar b=2$, the resulting $f(z)$ coincides with the results of the
simulation, as is shown in Fig. \ref{oscillate}(c).  
This clearly confirms our conjecture that 
there exists an oscillatory flow solution.

The configuration seen in Fig. \ref{oscillate}(c) can be regarded as 
the solution
that connects the two domains, namely, the right one filled with the
uniform flow solution, and the left one with the oscillatory flow
solution.
The transition region, or the domain wall, is moving at speed $V_0$.
Then, the boundary in the $\bar b-a$ plane
between the locally stable and the locally unstable region is understood
as the line where the speed $V_0$ of the domain wall connecting the two
solutions becomes zero;
$V_0<0$ ($V_0>0$) for the locally stable (unstable) parameter region.

%---------------figure 6--------------------------
\begin{figure}[t]
\begin{center}
\epsfig{file=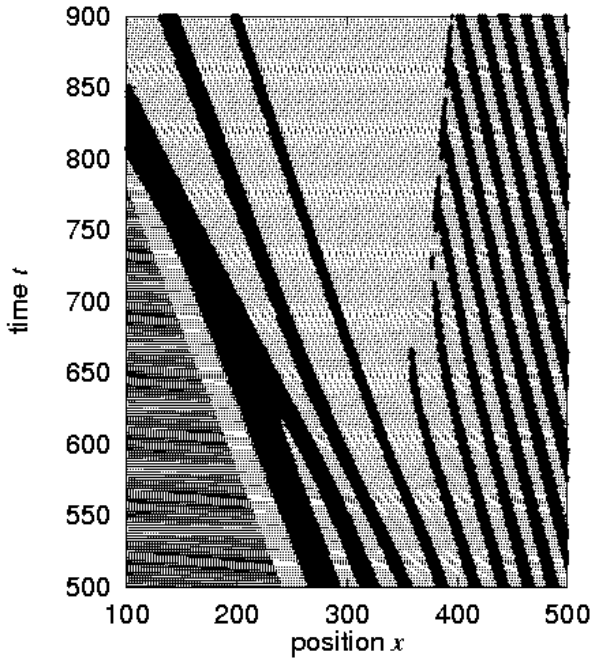, width=8cm,angle=-90}
\end{center}
\caption{
Spatiotemporal diagram  with $a=1.3$,
 $\bar{b}=2.5$, $L=1000$ and $\epsilon=0.1$. 
 The large headway region is seen to be invading the oscillatory flow 
region. 
}
\label{largeb}
\end{figure}
%-------------end of figure 6----------------
Near the point in the $\bar b-a$ plane where the local stability limit, 
defined by $V_0=0$, reaches the linear stability limit,
a new situation arises (Fig. \ref{largeb}).
The free-flow region with the lower car density, which emerges at 
the downstream end of
the alternating region, invades the oscillatory, and eventually, the
uniform flow.
The velocity $V_l$ by which the boundary between the oscillatory flow
and the low-density free-flow  advances
is determined by the continuity equation of the flux as
\begin{equation}
\frac{\bar{b}_l}{U(\bar{b}_l)-V_l}=\frac{\bar{b}}{U(\bar{b})-V_l},
\label{eq:V_l}
\end{equation}
where $\bar b_l$ is the headway of the low-density flow region.
Therefore, when $V_l>0$, the uniform solution is invaded by the 
low-density flow, even when $V_0<0$.
$V_l$ depends on $\bar b_l$, and $\bar{b}_l$ does not appear to be
determined solely by $a$.
It has, however, been numerically shown that 
$\bar{b}_l$ is always smaller than $\bar{b}_f$,
which is the headway of the free-flow under the PBC.
Thus, eq. (\ref{eq:V_l}) with $\bar{b}_l=\bar{b}_f$ yields
the stability limit,
which is also shown in Fig. \ref{phase}.

In summary, we have demonstrated that the uniform solution of the OV
model can be stable against a local perturbation, even in the linearly
unstable region if we employ the OBC.
We have also demonstrated that there exists an oscillatory flow solution in the
linearly unstable region and the local stability limit is determined
by the speed of the domain wall connecting the uniform flow solution
and the oscillatory flow solution.

Before concluding, let us make a few comments regarding the oscillatory
solution.
This is a new solution, which has not been noticed in the weak
nonlinear analysis near the linear stability limit, where eq. (\ref{eq:OVM}) is
reduced to the Korteweg-de Vries (KdV) or modified KdV equation.~\cite{KS}
The system chooses a particular phase speed $c_s$ in spite of the finite
continuous band of the phase speed being allowed.
This oscillatory solution has some characteristics in common with 
synchronized flow which has been observed recently 
in real traffic;~\cite{KR1,KR3,K}
that is the persistent fluctuating
flow excited by a localized perturbation in a high density region where 
the jammed phase can be nucleated.

%\newpage
%----------------references----------------------------


\begin{thebibliography}{99}
\bibitem{smalla}
The local stability limit cannot be determined in Fig. \ref{phase}
for the small $a$ and $\bar{b}$ region because the OV model is not valid
in these areas; the drivers' response is too slow and a car can 
pass ``through'' its preceding car in high-density regions.
\end{thebibliography}
\end{document}